# Content-Aware Tweet Location Inference using Quadtree Spatial Partitioning and Jaccard-Cosine Word Embedding

Oluwaseun Ajao
Sheffield Hallam University
United Kingdom
oajao@acm.org

Deepayan Bhowmik
University of Stirling
United Kingdom
d.bhowmik@ieee.org

Shahrzad Zargari
Sheffield Hallam University
United Kingdom
s.zargari@shu.ac.uk

*Abstract*—Inferring locations from user texts on social media platforms is a non-trivial and challenging problem relating to public safety. We propose a novel non-uniform grid-based approach for location inference from Twitter messages using Quadtree spatial partitions. The proposed algorithm uses natural language processing (NLP) for semantic understanding and incorporates Cosine similarity and Jaccard similarity measures for feature vector extraction and dimensionality reduction. We chose Twitter as our experimental social media platform due to its popularity and effectiveness for the dissemination of news and stories about recent events happening around the world. Our approach is the first of its kind to make location inference from tweets using Quadtree spatial partitions and NLP, in hybrid word-vector representations. The proposed algorithm achieved significant classification accuracy and outperformed state-of-the-art grid-based content-only location inference methods by up to 24% in correctly predicting tweet locations within a 161km radius and by 300km in median error distance on benchmark datasets.

## I. INTRODUCTION

The task of inferring user locations on Twitter as well as most social media platforms is non-trivial spurring the interests of many researchers in the field of artificial intelligence, computer science and behavioural sciences alike for almost a decade. Studies show that only less than 2% of Twitter users disclose or geotag the location of tweets [1] [2] due to fears of being tracked by online predators thus preserving their personal safety or by advertisers that use cookies to continually send them often times unsolicited product advertisements that have been personalised to their tracked location. Some social media sites even offer tailored location-based services such as Snapchat offering a new addition called SnapMap[1] where one can track the location of friends using the App and even know the status of their current activity including if they are sleeping or in ridding a car or shopping. These information are quite private and the users may not even be aware they have provided such information which could lead to stalking and posing threats especially for children [3]. However location tracking of the online users also has benefits relating public safety and security.

The growing threat of online crimes ranging from messages focused at propagating hatred, to cyberbullying and spread of fake news and false information for the purpose of promoting malicious selfish intentions; personal or political gains have continued to cause governments, corporate organisations and individuals cause for concern. Social media is a good tool for the promotion of information but the fact that it is uncensored - stemming from the notion of freedom of speech which obtains in most democracies tend to be abused. It is crucial that law enforcement bodies are able to track down the location of these offenders and the origin of these messages to curtail their spread before they begin to 'infect' the behaviours and actions of other online users.

The large footprint of Twitter makes it an important marketplace for advertisers to reach their consumers, and serve as projection platforms for the government to its citizens. Knowledge of users who interact on Twitter may be quite useful for organisations that render these services. There exist third party domains and other sources such as knowledge bases. These sources amongst others are useful for estimating user locations [4]. However, they may be unreliable and insufficient for effectively estimating the location of users. This brings the need to infer locations from transmitted messages solely based on the content alongside other relevant metadata information captured with the tweets such as user description and time zone information etc.

In this paper we propose a novel non-uniform grid-based approach for location inference from Twitter messages combining quadtree spatial partitions and semantic understanding using Natural Language Processing (NLP). The contributions made in this work is given as follows.

- A discriminative grid-based approach for the determination of tweet locations based on the content,
- A Quadtree spatial indexing technique for inferring locations based on variable nodes,
- A NLP based hybrid word embedding model consisting of Cosine and Jaccard similarity measures [5] for dimensionality reduction in the feature vector, and representation [6].

---
[1] www.snapchat.com

- Improvement in grid-based location inference based solely on the content of Twitter messages.

A functional block diagram of the proposed algorithm is depicted in Fig. 1. In the remainder of this paper Section II gives related works in the field of location inference and NLP, Section III gives description of the methodology used during the experiment. Analysis and results are presented in Section IV while Section V gives the conclusions.

## II. RELATED WORKS

Location inference also referred to in literature as *Geolocation Prediction* has enjoyed a fair amount of research interests by several authors working within the space. A few works have been written on the inference of location of Twitter users. The one most related to this work is the [7] where the authors estimated user locations solely based on the content of their messages using supervised classification. The work extracted *local words* from the messages of users with the assumption that users from specific geographic locations would normally use words that are *local* to that geographic location. For example the word *howdy* which is *hello* in English would be considered to be more frequently used in the US state of Texas. However, the authors did not actively seek out to recognise entities such the names of people, places and organisation within twitter messages as part of their location inference technique unlike our proposal in this work. It should be noted that some of such local words they identified could also be geo-entities, for instance their probabilistic method identified the word *ucsb* to show a peak distribution around the state of California as this was the abbreviation for the University of California located in the city of Santa Barbara.

Location inference and privacy of geo-spatial data have always been an area of concern. Krumm [21] examined the identification of users from web search data and was able to successfully identify their locations to the granularity level of home addresses from GPS data. This is was possible due to the very high degree of correctness that GPS data typically offers. However, the availability of location information is not always guaranteed which introduce additional challenges. Our approach aims to address this issue by inferring the user locations to a city-level accuracy by analysing user texts available from social media. Privacy continues to be an emerging area of research discussion [20] with people choosing to hide their online identities to keep an anonymous profile from other users and in some cases for the safety and the fear of being *trolled* online by cyberbullies especially in the social media and Twitter in particular.

Han *et al.* [22] used words referred as *Location Indicative Words* (LIWs) and provided a spatial clue to indicate the whereabouts of the users. It was proposed that users were more likely to be successful in preserving their privacy if they refrain from mentioning these LIWs in their online conversations and also to actively delete location meta data from their online footprint. This seems far from being realistic as users will most likely be tracked by the social media platforms who passively collect and retain time-stamped information such as time-zones and IP addresses of their users, Most of these meta data is then made available to the public via the Twitter API and can be linked it to the users who created them. Other works done in the field can be found in [11] that proposed a method which learns association from locations and keywords from previous user messages to predict subsequent messages. The challenge with this method is that to effectively train a location classifier the past tweets of a user would have to be collected and analysed and may be prove to be technically infeasible because at the moment the Twitter REST API only allows the retrieval of the last 3200 messages of any user. Secondly there is the possibility that users can relocate over time from one city to another or even from one country and time zone to another. Thus online themes and conversations that they tweet about today may be different tomorrow. Our approach is not user-specific and relies on word-usage and geo-entities associated with locations.

Jurgens [23] applied label propagation of location assignments to the knowledge of locations. The work relied on the friends connections also known as their *ego* network locations including self-reported ones found in the free-text fields of the user profiles. Compton *et al.* [15] inferred location from the friends network with known locations. Their work presented the largest dataset utilised till date for the training and testing of their location inference classifier accounting for tweets captured from over 100 million Twitter users. Chang *et al.* [12] used Gaussian mixture models and maximum likelihood estimation (MLE) which is purely content-based in addition to the use of local words distribution within messages. Mahmud *et al.* [16] used an ensemble of statistical and heuristic classifiers. Their approach also followed a hybrid of both tweet content and social network profile information including the friends networks. Ajao *et al.* [4] gave an insight into a range of clues for estimating user locations in addition to the message content. They outlined three various locations that had been inferred in on Twitter including tweeting location, user home residence and message context that have mentions or references to certain geographical locations or points of interests. Various partitioning algorithms are proposed in the literature to infer Tweet locations including k-dimensional trees [13], [17] or uniform grids [9], [19]. A further breakdown of reported results from related works is presented in Table I.

We believe the task of location inference from tweets and other sources that involves the use of text, relies on natural language processing models and machine learning techniques to understand the semantics. There are over 500 million messages sent by Twitter users each day[2]. Thus, it is humanly impossible to manually sift through the contents of these messages and make meaning of them. NLP models such as word embedding and pattern recognition capabilities of machine learning models are useful in the identification of patterns [24] within the text. This helps in machine understanding of the human language and drawing insights suitable for the process. NLP methods applied in this work includes the use of the continuous bag of word (CBOW) model [25] for embedding the words into vectors. Additionally, Jaccard similarity and Cosine distance of word vectors [5], [26] was computed for feature extraction and word dimensionality reduction to get prediction-relevant text. To the best of our knowledge our approach is the first to use Quadtree spatial indexes combining with NLP for content-aware location prediction on Twitter.

[2]www.twitter.com

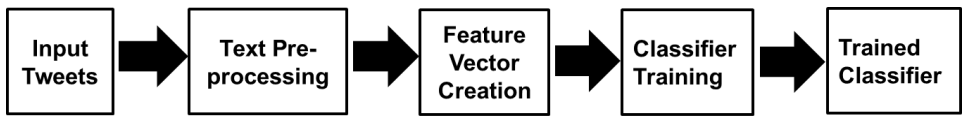

Fig. 1. Functional diagram illustrating the tweets location inference task.

TABLE I
METHODS AND OUTCOMES FROM RELATED WORKS IN TWITTER LOCATION INFERENCE

| Author | Input | Method | Technique | ACC(%) | Radius |
|---|---|---|---|---|---|
| Cheng *et al.* [7] | content | words | Probabilistic(ML) | 51 | 160km |
| Eisenstein *et al.* [8] | content | geo-topic | Geo-Topic Model | 24 | State |
| Wing *et al.* [9] | content | locations | Grid-based(Uniform) | - | - |
| Kinsella *et al.* [10] | content | locations | Language Models | 13.9 | Zip Code |
| Kinsella *et al.* [10] | content | locations | Language Models | 29.8 | Town |
| Ikawa *et al.* [11] | content | words | ML classification | 20-60 | 10-30km |
| Chang *et al.* [12] | content | words | GMM & MLE | 49.9 | 160km |
| Roller *et al.* [13] | content | locations | Grid-based(kd-tree) | 34.6 | 160km |
| Li et al [2] | content, network | hybrid | Probabilistic(ML) | 66 | 160km |
| Schulz *et al.* [14] | content, context | hybrid | Gazetteer | - | - |
| Compton *et al.* [15] | Network | closeness | Network | 80 | |
| Mahmud *et al.* [16] | content, context | locations | Ensemble classifiers | 58 | city-level |
| Wing *et al.* [17] | content | locations | Grid-based(kd-tree) | 90.2 | 160km |
| Ryoo & Moon [18] | content | words | Location services | 56.7 | 10km |
| Hulden *et al.* [19] | content | words | Grid-based(Uniform) | - | - |
| Han *et al.* [20] | content | words | Neural Net | 40.9 | - |

### III. METHODOLOGY

We propose a new grid-based approach for location inference from Twitter messages using quadtree spatial partitions. The proposed algorithm incorporates Cosine similarity and Jaccard similarity measures for NLP-based feature vector extraction and dimensionality reduction. The summary of the illustration of our approach towards content-based location inference is given in Fig. 1 and described in detail in this section.

#### A. Feature Vector Creation using NLP

The proposed work incorporates natural language processing methods in creating the feature vector. This includes word embedding, where the words are converted to numbers in order to process them effectively and forming the vector representation of those words. There are two broad categories of word embedding namely, *a)* frequency-based and *b)* prediction-based word embeddings as seen in Fig. 2. Available models for word embedding include the continuous bag-of-words CBOW model [27], Word2vec model [25] and Glove model [28]. Prediction based word embedding techniques such as *word2vec* are proved to be the state-of-the-art technique and have the advantages over deterministic methods such as conventional bag-of-words or count vectors. These also have the ability to incorporate neural networks improves their performance compared to their predecessors. For our work we adopted the *word2vec* model proposed in [25].

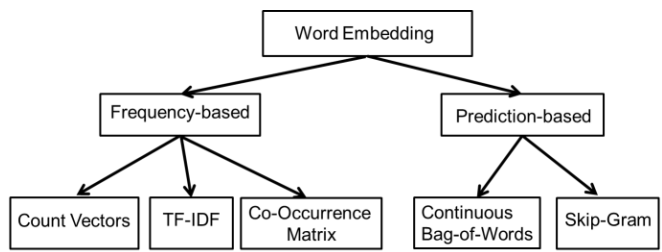

Fig. 2. Methods of Word Embedding in Natural Language Processing

Our NLP based text processing for feature vector creation includes following three central steps:

**Calculation of linear vector:** Linear vector calculations are implemented on feature vectors using *word2vec*. An example of this is *King - Man + Woman = Queen*. This is inherent in the fact that once words are converted into vectors they lend themselves to algebraic and mathematical operations thus revealing the association and relationships that exist between them. In the above example the gender is the relationship between them.

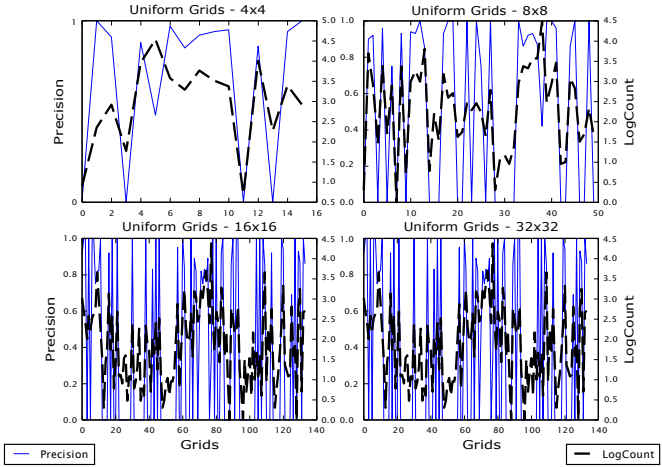

Fig. 3. Interaction between precision and log of uniform grid counts

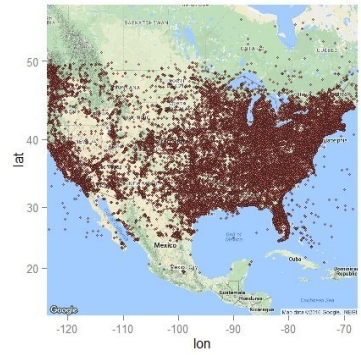

Fig. 4. Geo-located US Tweets partitioned with our Quadtree algorithm

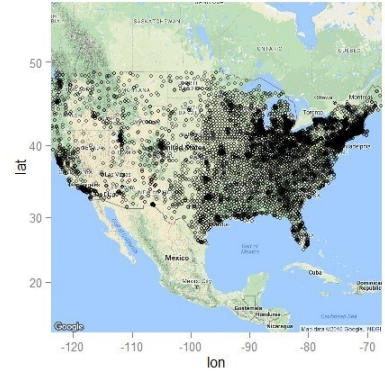

Fig. 5. US cities with population over 5,000.

- **Identification of synonyms used in the messages:** Words that have the same semantic meaning are given the same representation. In essence it looks out for word synonyms avoids redundancy and significantly reduces the size of the feature vector and computing time. For example the two sentences $S_1$ = {*That is a small thing*} and $S_2$ = {*That is a little thing*} will be considered the same, thus improving the effectiveness of their respective word-vector representation.
- **Determination of similarity threshold:** Similarity thresholds can be specified where the distance between the feature vectors is measured with the cosine similarity functions (described in Eq. (1) and Eq. (2)). In this regard, words that have similar syntactic appearance but were however mis-spelt, exaggerated or abbreviated would be recognised and given the same representation within the vector space. This can be achieved by the cosine function to compare their similarity with the English language dictionary. For example, *Yeeeees* is equivalent to *Yes* or *Gooooood* is recognised as *Good*.

In order to measure the closeness between the word feature vectors we have used two types of similarity measures [5]: *a) Cosine similarity* and *b) Jaccard similarity* as described below. Considering two non-zero vectors, $p$ and $q$, each having component values $1, 2, ...n$, their cosine similarity ($Sim_c(p, q)$) can be calculated as:

$$Sim_c(p, q) = \frac{p.q}{\|p\|_2.\|q\|_2} = \frac{\sum_{i=1}^{n} p_i.q_i}{\sqrt{\sum_{i=1}^{n} p_i^2}.\sqrt{\sum_{i=1}^{n} q_i^2}}. \quad (1)$$

For the same vectors the Jaccard similarity ($Sim_j(p, q)$) is calculated by:

$$Sim_j(p, q) = \frac{|p \cap q|}{|p \cup q|}. \quad (2)$$

### B. Sparsity of Tweets and Quadtree

Tweeting geo-locations bear close resemblance to actual population demographies, showing similar density patterns. Comparing Fig. 4 and Fig. 5 and considering that the continental United States has a total geographic area of 6,110,264 square miles [29], we observed this challenge with the dataset is due to the geographical outlay of the country as some areas were more inhabited than others. Thus tweets were considered to have a sparse distribution in some area. As such when the map was uniformly split some contained too little while some contained too much. This presented a major limitation in the estimation of location of the users using a uniform grid method. In contrast, Quadtrees being hierarchical spatial data structures [30] offers a solution that can dynamically address such issue. Fig. 3 shows the correlation between the number of tweets within each of the grids namely 4x4, 8x8, 16x16 and 32x32 all done using the uniform spatial partitioning method. The precision value was plot against each of the 4 split grids. We found a direct relationship between the log-value of the counts of observations each uniformly partitioned grid and their precision. This implies the presence of a bias favouring highly populated grids while the less populated ones got lower precision. Following this observation, we chose to cluster the dataset in a discriminatory manner now dependent on the counts of observations within each grid. This created more effective labeled training dataset for the classifiers.

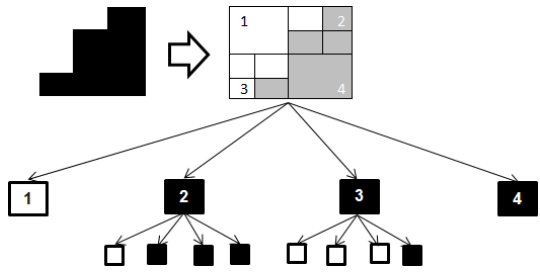

Fig. 6. Spatial partitioning method illustrating decomposition and equivalent Quadtree representation of an object

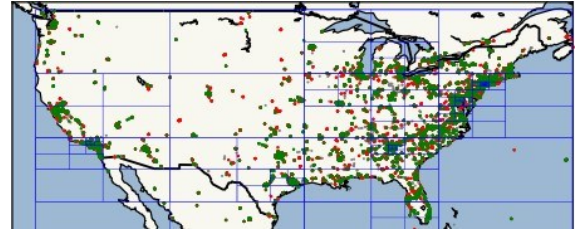

Fig. 7. Geo-located US Tweets partitioned with our Quadtree algorithm

*1) Quadtree Data Partitioning:* Quadtree is a hierarchical data structure and partitioning technique for efficiently organizes data in a pre-defined discriminative manner. As a result, we find it suitable in addressing the bias mentioned earlier in this section. Quadtree illustration is given in Fig. 6. The object is first decomposed into four quadrants or nodes; numbered 1,2,3 and 4. Nodes 2 and 3 are further split based on spatial interest into 4 leaves each. Quadtrees effectively handle spatial querying of geographic data [31] and proven applications in the areas of collision detection and image processing [30]. We employ the method because in the area of location inference application of the algorithm would be to query the location of users around a predefined radius or some location of interest such as a geo-political region. This method fulfills this requirement and as such within a conventional database ultimately support searches, insertions and deletions within the parent and leaf nodes of the dataset.

Another advantage that the Quadtree method has over uniform grids is also the time saved in implementation as the nodes with little or no data points can be easily dropped from the query. Our implementation considered a variable resolution constraint which can be adjusted. For empirical purposes and during the decomposition process we set the maximum number of points in each grid in multiples of 5,000 or lesser, *i.e.*, 10000, 15000, 20000 etc. As the dataset is now more fairly split across more grids we observe significant correlation between the log values of the grid counts as well as improvement in the level of accuracy, AED, MED further described in Section IV.

*2) Tweet based bias removal:* In addition to the Quadtree partitioning we also use a population based bias removal technique. The geo-spatial visualisation of the US tweets (refer Fig. 4) indicates more visible activities towards the North East of the country; this bears a true resemblance of Fig. 5 which illustrates the population of the United States [32] as discussed earlier. This implies that there is a bias favoring a larger count size as opposed to less dense grids. This is clearly a problem due to the sparse distribution of the tweets and as we see from the geographical map, tweets on the East coast (around New York etc tend to have normally a larger population density and thus more user tweets are included in the training data for this purpose). In order to further remove this bias we used a weighted measurements of the outcome and incorporated this within the measurement metric as discussed in Section IV. Fig. 7 was generated using our Quadtree structure.

### C. Training of Location Classifier

The task of content-based location inference can be interpreted a classification problem. A number of machine learning classifiers were examined including the Logistic Regression (LOGIT) / Maximum Entropy, Random Forests, Decision trees, Artificial Neural Networks and the Multinomial Naive Bayes (MNB) classifier for supervised classification of more than 730,000 messages geotagged to the continental United States. Preliminary investigation with baseline datasets namely GEOTEXT [8] and UTGEO-Small [13] indicates better performances by Multinomial Naive Bayes and Logistic Regression which are also commonly used in similar dataset by other researchers. This subsection briefly revisited these two classifier before reporting the results. In the classification, the words served as the features while the grids served as the labels or predicted results of the task. In training and testing the classifier, we implemented a ten-fold cross validation technique for training both LOGIT and MNB models. In line with Twitter's privacy policy, the dataset used is anonymised. user names are discreetly represented as randomly assigned numbers. During the pre-processing stages of our training dataset, we removed duplicate messages. However, it should be noted that chat bots may not always generate duplicate messages. In such instance, there may be a bias towards such locations. We leave the review of this impact to future studies.

*1) Multinomial Nave Bayes:* This is quite popular with discrete probability distributions such as word counts in text classification. They are straight forward to implement. We set an alpha value at 1 and learn class prior probabilities; they were also adjusted according to the classes within the dataset. With a multinomial event model, samples (feature vectors) represent the frequencies with which certain events have been generated by a multinomial $(p_1, ..., p_n)$ where $p_i$ is the probability that event $i$ occurs (or K such multinomials in the multiclass case). A feature vector $X = (x_1, ..., x_n)$ is then a histogram $x_i$, with counting the number of times event i was observed in a particular instance. This is the event model typically used for document classification, with events representing the occurrence of a word in a single document (see bag of words assumption). The likelihood of observing a histogram $x$ is given by:

$$p(x|C_k) = \frac{(\sum_i x_i)!}{\prod_i x_i!} \prod_i p_{ki}^{x_i} \quad (3)$$

*2) Logistic Regression:* Also known as the Maximum Entropy or Logit classifier is a regression model where the dependent variable(s) are categorical; In the generalized additive form consider a set of independent variables $X_1, X_2, ..., X_p$ predicting a likely outcome (Y) with $f_j (f_1, f_2, ..., f_p)$ unspecified smooth functions. The model is given as:

$$\log \frac{\mu(X)}{1 - \mu(X)} = \alpha + f_1(X_1) + f_2(X_2) + ... + f_p(X_p) \quad (4)$$

In our experiment the Logit classifier outperformed all the classifiers in terms of precision, accuracy and recall. We use the L2-penalty as it is more robust and handles sparse data well; A phenomena that is quite common with geo-located Twitter data sets. We did not implement dual or primal formulation as the number of samples far exceeded the number of the features in the task. We adopt a balanced class scaling applied to handle the L2 penalty. We set the maximum number of iterations at 100.

## IV. RESULTS AND DISCUSSIONS

This section describes the measurement metrics that were used for evaluation of our proposed method, the results and related discussions.

### A. Data

Two datasets were used in training our MNB and LOGIT classifiers and evaluation of our technique. These include the UTGEO-Small Dataset [13]. This consists of 670,000 geolocated Twitter messages from the continental United States. The GEOTEXT dataset by [8] was also used. This had approximately 380,0000 geotagged US tweets.

### B. Data Analysis: Measuring Classifier Efficiency

After the texts were cleaned the resultant vector was significantly reduced. Afterwards matrices were parsed into word vectors before the machine learning task was performed on them. Given a set of geo-tagged tweets $T_i = \{t_1, t_2, ..., t_N\}$ with ground truth location of $ActualLoc_i$ and a predicted location $ExpectedLoc_i$. We evaluate the classifiers performance using the Average Error Distance (AED), Median Error Distance (MED) and Distance-Based Accuracy.

*1) Calculating Error Distance:* In calculating the error distance (in km) between the predicted and actual location we apply the Haversine formula [33], [34] also known as the *Great Circle Distance* between any two geo-coordinates on the earth's surface assuming an spherical shape of the Earth. This method was chosen as we found suitable and stable in determining the distance estimation of several diverse and closely located geo-coordinate pairs.

*2) Average Error Distance:* The average error distance (AED) measures the arithmetic average error of predictions for the messages within the dataset, however it should be noted that this metric can be easily skewed by large ranges of values within the dataset and would particularly be unreliable if there where a presence of an anomaly in the training of the classifier. The AED for the classifier is given thus:

$$AED = \frac{1}{|N|} \sum_{i=1}^{N} \{|ActualLoc_i - ExpectedLoc_i|\}. \quad (5)$$

*3) Median Error Distance:* The median error distance (MED) overcomes the limitation of the AED by considering only the error values close to the median. We find this to be most reliable and gives a truer indication of the performance. This is represented as:

$$MED = MedianDistance\{|ActualLoc_i, ExpectedLoc_i|\}. \quad (6)$$

*4) Distance-Based Accuracy:* Accuracy levels at a set distance and more specifically around a distance d is a renowned benchmark and this was applied in this work also. It is estimated as the ratio of correctly predicted location with an error margin less than $d = 161$km compared to the entire tweets count

$$ACC@161 = \frac{\{|ActualLoc_i - ExpectedLoc_i|\} \leq 161km}{|N|}. \quad (7)$$

### C. Experimental Results and Discussion

The summary performance of our method, measured against various metrics such as average error distance (AED) as calculated in Eq. (5), median error distance (MED) and predicted accuracy to the nearest 161km from Eq. (6) and Eq. (7) respectively are shown in Table IV. A detailed breakdown of each of the classifiers (MNB) and (Logit) for both GEOTEXT datasets is given in Table II while that of the UTGEO-Small dataset is given in Table III.

On the GEOTEXT dataset from Table II we achieved significant improvements when the grid counts are reduced from 20,000 all the way to 5,000 tweets in all three metrics specifically MED of 39.15km, AED of 598.44km and 59.47km performing better than methods that had been applied on the same dataset [8], [9], performing better than [19] by more than 150km in AED and almost 300km in MED Similarly, from Table III and IV our method performed better than [13] by 24% in terms of ACC@161, more 400km better prediction for median error distance and over 250km more accurate average error distance. This implies that we are unable to go beyond the maximum grid size and a granularity level finer than 5000 tweets as this could be lead to overfitting of the training data.

TABLE II
QUADTREE-BASED CLASSIFICATION SHOWING ERROR DISTANCE AND COMPUTE TIME FOR TWO DIFFERENT CLASSIFIERS ON GEOTEXT DATASET

| Grid Count | Med-ED (km) | Avg-ED (km) | ACC@161 | Time (mins) |
|---|---|---|---|---|
| Logit - GeoText dataset | | | | |
| 20,000 | 143.98 | 571.39 | 51.72 | 68 |
| 15,000 | 125.73 | 700.76 | 53.84 | 70 |
| 10,000 | 129.18 | 620.81 | 52.04 | 73 |
| 5,000 | 39.15 | 598.44 | 59.47 | 79 |
| MNB - GeoText dataset | | | | |
| 20,000 | 411.22 | 721.11 | 38.57 | 58 |
| 15,000 | 1009.82 | 579.44 | 41.61 | 58 |
| 10,000 | 279.78 | 876.67 | 30.76 | 58 |
| 5,000 | 400.62 | 853.33 | 42.57 | 58 |

TABLE III
QUADTREE-BASED CLASSIFICATION SHOWING ERROR DISTANCE AND COMPUTE TIME FOR TWO DIFFERENT CLASSIFIERS ON UTGEO-SMALL DATASET

| Grid Count | Med-ED (km) | Avg-ED (km) | ACC@161 | Time (mins) |
|---|---|---|---|---|
| Logit - UTGeo-small dataset | | | | |
| 20,000 | 249.45 | 833.10 | 43.70 | 78 |
| 15,000 | 124.44 | 651.81 | 54.75 | 78 |
| 10,000 | 92.86 | 618.07 | 57.30 | 79 |
| 5,000 | 45.00 | 600.79 | 60.24 | 81 |
| MNB - UTGeo-small dataset | | | | |
| 20,000 | 665.31 | 1093.45 | 30.20 | 71 |
| 15,000 | 449.78 | 907.65 | 40.07 | 71 |
| 10,000 | 418.08 | 828.13 | 42.56 | 71 |
| 5,000 | 380.76 | 855.64 | 43.76 | 71 |

TABLE IV
OUR METHOD AND OTHER GRID-BASED RESULTS

| Author | Method | AED | MED | ACC@161 |
|---|---|---|---|---|
| GeoText dataset | | | | |
| Eisenstein et al. [8] | Topic Models | 900 | 494 | 24 |
| Wing et al. [9] | Uniform | 967 | 479 | N/A |
| Hulden et al. [19] | Uniform | 764.8 | 357.2 | N/A |
| Our | Quadtree | **598.44** | **39.15** | **59.47** |
| UTGeo-small dataset | | | | |
| Roller et al. [13] | kd-tree | 860.0 | 463.0 | 34.6 |
| Our | Quadtree | **600.79** | **45.00** | **60.24** |

TABLE V
ERROR DISTANCE IN MILES FOR ACC@90% BY OUR LOGIT CLASSIFIERS

| | Logit-20k | Logit-15k | Logit-10k | Logit-5k |
|---|---|---|---|---|
| GEOTEXT | 1,046.28 | 1,733.03 | 1,237.38 | 1,193.57 |
| UTGEO - Small | 1,558.73 | 1,253.70 | 1,230.40 | 1,289.83 |

In terms of the computing time to execute both methods using our algorithm we see from both datasets the MNB was quicker to execute than the Logit classifier. On average about 20 minutes longer to execute. While our method performs better on all metrics namely AED, MED and ACC@ 161, it should be noted that the AED is less influenced by anomalous values in the training dataset as it relies on median values. However, MED should be given more consideration over the AED as the latter can be affected by a range of very low and very high error distances thus not giving a fair assessment of the classifier performance.

We have also shown the performances of the method with and without considering demographical biases as discussed in III-B2. It is evident the performance has improved significantly while we removed the bias using a weighting parameter that is proportional to the demographic distribution. Finally we compare our method with other grid-based methods in Table IV. The result indicates that our method outperformed the existing grid-based location inference techniques on Twitter. Showing significantly better results in terms of AED, MED and Accuracy at an error distance of 161km radius. Wrong predictions are also included in the visualisation in Fig. 7

According to the 2017 US Census [35], the average size of a city with a population above 100,000 is approximately 100 sq miles. The average land area of a US state is over 57,000 sq miles. While an average county size is 1,124 sq miles. At an error distance of just over 1,100 miles we achieve 90% accuracy as shown in Table V. Thus, we place the performance of our Logit classifier quite competitive with the state-of-the-art at the city-level granularity and much more so at the county level. It would be interesting to know how our results compare against deep learning text classification methods, if there's significant improvement in terms of accuracy, AED and MED. It is currently out of the scope of this study.

V. CONCLUSIONS AND FUTURE WORK

This paper proposes a new non-uniform Quadtree grid-based approach for location inference from Twitter messages. The proposed algorithm uses natural language processing for semantic understanding and incorporates Cosine similarity and Jaccard similarity measures for feature vector extraction and dimensionality reduction. The result of the grid classification shows good improvement over the existing state-of-the-art grid based approaches in city-level location inference on existing benchmark dataset of the GEOTEXT and UTGEO-Small Twitter corpuses. 60% of tweets are accurately predicted within an error distance of 161km (100 miles radius). The results show the effectiveness of our Quadtree spatial indexing technique in combination with a Logistic regression classification model which outperforms other grid-based methods in location inference. Future work could look the location prediction in real-time from live Twitter data streams and possibly linking with other location-based networks and for other geographical regions of the world. There is also a potential application in helping to address online social media issues such as fake news detection [36] and tracking the origin of online malicious content-based messages.